\documentstyle[11pt,aaspp4,psfig]{article}
\lefthead{Whiting, Irwin and Hau}
\righthead{Antlia Dwarf Galaxy}

\begin{document}
\title{A New Galaxy in the Local Group: the Antlia Dwarf Galaxy}

\author{Alan B. Whiting}
\affil{Institute of Astronomy, University of Cambridge}
\authoraddr{Madingley Road, Cambridge CB3 0HA, UK}
\author{M. J. Irwin}
\affil{Royal Greenwich Observatory}
\authoraddr{Madingley Road, Cambridge CB3 0DZ}
\and
\author{George K. T. Hau}
\affil{Institute of Astronomy, University of Cambridge}
\authoraddr{Madingley Road, Cambridge CB3 0HA, UK}

%\simlt and \simgt produce > and < signs with twiddle underneath
\def\spose#1{\hbox to 0pt{#1\hss}}
\def\simlt{\mathrel{\spose{\lower 3pt\hbox{$\mathchar"218$}}
     \raise 2.0pt\hbox{$\mathchar"13C$}}}
\def\simgt{\mathrel{\spose{\lower 3pt\hbox{$\mathchar"218$}}
     \raise 2.0pt\hbox{$\mathchar"13E$}}}
\def\Msun{{\rm\,M_\odot}}

\begin{abstract}
We report the discovery of new member of the Local Group in the
constellation of Antlia. Optically the system appears to be a typical 
dwarf spheroidal galaxy of type dE3.5 with no apparent young blue stars or 
unusual features. A color-magnitude diagram in $I$, $V-I$ shows the tip of
the red giant branch, giving a distance modulus of 25.3 $\pm$ 0.2
(1.15 Mpc $\pm$ 0.1) and a metallicity of -1.6 $\pm$ 0.3.  Although Antlia is 
in a relatively isolated part of the Local Group it is
only 1.2 degrees away on the sky from the Local Group dwarf NGC3109,
and may be an associated system.

\end{abstract}

\section{Introduction}

Despite their unassuming appearance dwarf galaxies hold the key to 
many questions of galaxy formation, structure and evolution. 
Their spatial distribution and the
information they provide about the faint end of the galaxy luminosity 
function provide key constraints on theories of galaxy formation, while
the input they provide to theories on the nature of dark matter and
of star formation in low density enviroments are crucial to our understanding
of these issues (see, for example,
Binggeli, Sandage, \& Tamman \markcite{bst88} 1988).  
The internal dynamics of dwarf galaxies has given evidence of an unusually high
ratio of dark to luminous matter (Mateo \markcite{m94} 1994, Hargreaves et 
al.\ \markcite{hgic96} 1996) and the complexity of the star formation history 
of the nearby Local Group dwarf spheroidal (dSph) galaxies continues
to challenge theories of star formation (Hodge \markcite{h94} 1994, 
Elmegreen et al.\ \markcite{eesm96} 1996).  Within the Local Group
the motions of the more distant dwarf members can be used to 
probe both the dark matter halo of the Group as a whole and via the 
timing arguement provide an independent estimate of the age of
the Universe (Lynden-Bell \markcite{lb81} 1981). 
For all of these studies more examples of nearby dwarf 
galaxies are needed.  

\begin{figure}
\centerline{
\hbox{\psfig{figure=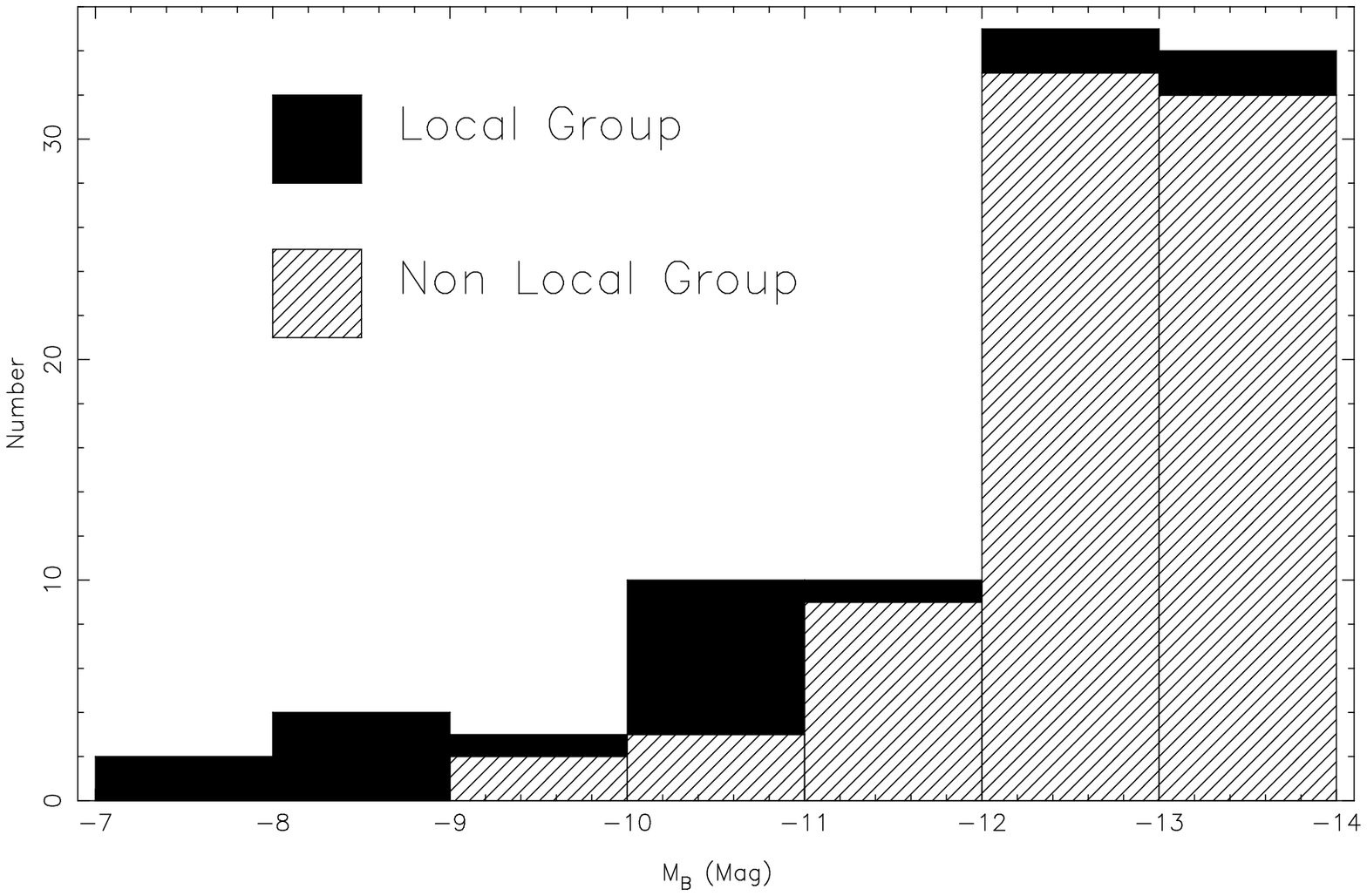,width=15cm}}
}
\label{lf}
\end{figure}

However, by their nature dwarf galaxies are elusive and comparatively few are 
known, especially extreme dwarves  with M$_{\rm B} \simgt -11$. In a sample of
nearby galaxies intended to be complete to $10$ Mpc (Schmidt \& Boller
\markcite{sb92} 1992), essentially all galaxies with $M_{\rm B} > -11$
are in the Local Group.  Figure~\ref{lf} illustrates how the faint end 
of the luminosity function of
nearby galaxies is dominated by the contribution of the Local Group members.
This indicates that extreme dwarves must be
sought nearby. Searches have been undertaken toward M31 and three
local dwarves found (Van den Bergh \markcite{vdb72} 1972).  An
automated search in high galactic latitudes found the Sextans dwarf
(Irwin et al.\ \markcite{ibbm90} 1990).  However, Local Group dwarves,
especially those on the outskirts, may be found anywhere on the sky,
so an all-sky search is necessary to find them.

To this end a visual inspection of glass copies of all 894 ESO-SRC IIIaJ 
survey plates covering the southern sky ($\delta < 3^\circ$) has been 
performed.  Objects resembling the Andromeda dSphs
and the Tucana dwarf, that is of very low surface brightness (VLSB), diffuse 
and large ($\simgt 1'$), were noted and agumented by similar objects picked up 
during the UK Schmidt Telescope Unit (UKSTU) visual inspection of the original 
survey plates (S. Tritton, personal communication).  The astronomical 
reality of the candidates was checked, where possible, by examining the 
equivalent ESO R survey glass copies; the extra color information also
providing a useful discriminant against Galactic reflection nebulosity.
After cross-checking our list of VLSB objects with published catalogues of 
extragalactic objects having known radial velocities and against catalogues
of known Galactic planetary nebulae - the main ``contaminant'' at low Galactic 
latitudes - the remaining 75 candidates were digitized on the PDS 
microdensitometer at the Royal Greenwich Observatory.  On the digitized scans 
several of these candidates appeared to be marginally resolved at the limiting
magnitude, B$_{\rm J} \approx 22.5$, of the photographic survey material.  
Deeper followup imaging of the candidates revealed one of them to be a 
previously unknown Local Group galaxy in the constellation of Antlia.

\section{Followup Observations}

Objects from the VLSB list were imaged using the 1.5m telescope at
Cerro Tololo Interamerican Observatory during the period 2 -- 6 March
1997 using a thinned Tek 2048 $\times$ 2048 CCD as detector.  At the
f/13.5 Cassegrain focus of the 1.5m this results in a scale of 0.24 
arcsec per pixel and a field coverage of just over 8 $\times$ 8 arcmin.
Candidates were initially examined by taking 20 minute exposures
in the R-band.  With the seeing typically between 0.9 -- 1.4 arcsec this
enabled stellar objects to R $\approx$ 23 to be detected.  At this depth
objects close to or within the Local Group should begin to resolve into
stars, with the tip of the Giant Branch becoming readily visible.  
If a candidate appeared to resolve into stellar components further broadband
observations in the V and I-bands together with narrowband H$\alpha$ were 
obtained.   The raw CCD frames were processed in the standard way 
(bias-subtracted, trimmed and flat-fielded using twilight flats taken during
the observing run) in almost real time at the telescope to aid in visual
inspection of candidates.  A series of standard star fields taken from the
list of Landolt \markcite{la92} (1992) were observed at 
intervals throughout each night.
Conditions were generally photometric with the seeing very stable and averaging
1.2 arcsec.

\begin{figure}
\centerline{
\hbox{\psfig{figure=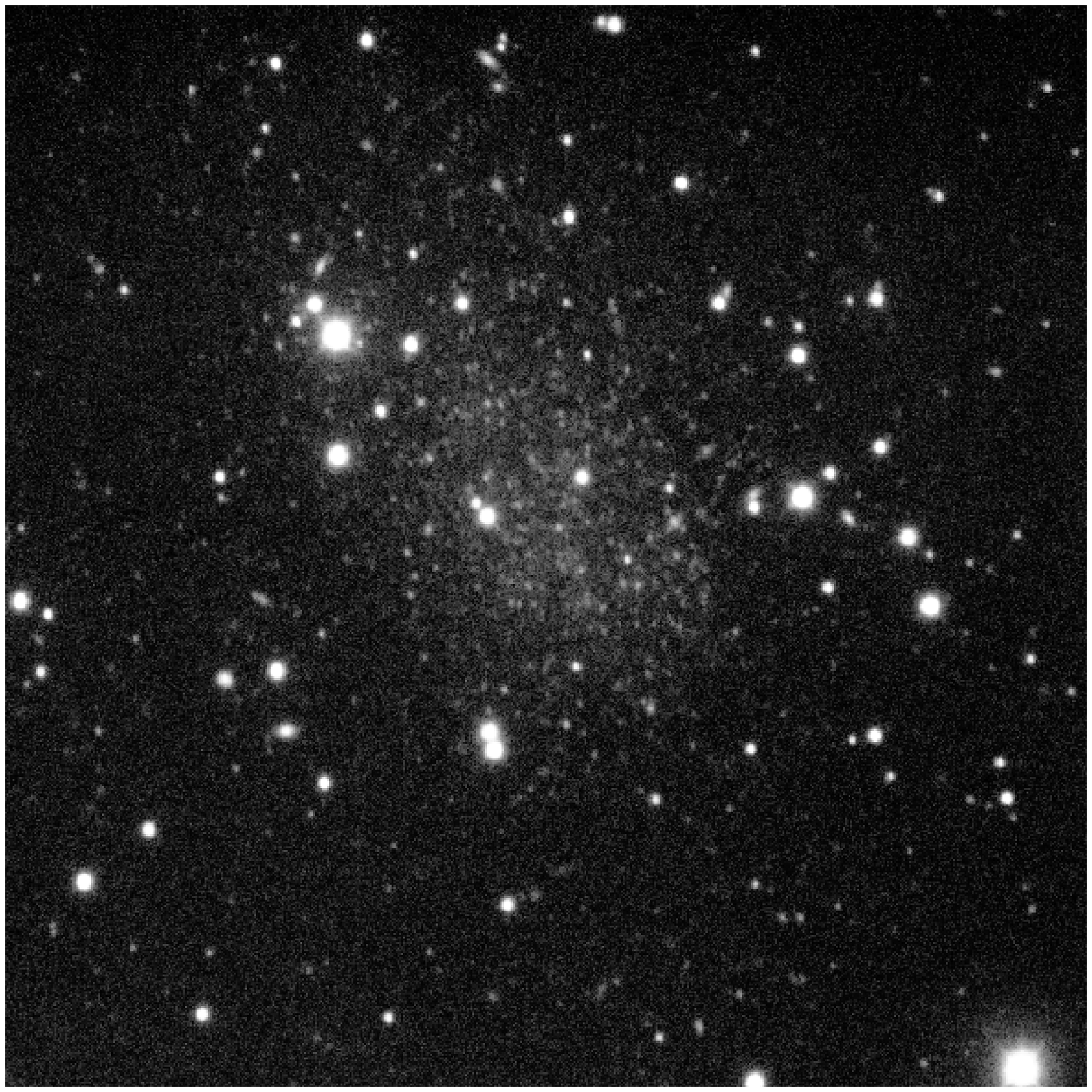,width=12cm}}
}
\label{antlia}
\end{figure}

An initial exposure of 20 minutes in R showed the Antlia dwarf, located
at $\alpha = 10^h 01^m 17^s.5$, $\delta = -27^\circ 05' 15''$ (B1950),
to clearly resolve into stars.  Further observations totalling 4800s in V,
3600s in R and 3600s in I were subsequently made to explore the nature of the 
galaxy.  An 1800s H$\alpha$ image 
revealed no obvious concentrations
of young stars or regions of ionised gas.  The individual broadband frames
in each passband were coaligned and combined to eliminate the effect of
cosmic rays.  A ``true'' color picture derived from 
the combined V,R,I images
of the Antlia dwarf is shown in figure~\ref{antlia} and clearly reveals
the brightest stellar component of the galaxy. 

\section{Analysis}

\subsection{Morphology and magnitude}

The distribution of stars revealed by the color image in Figure~\ref{antlia}
shows no obvious concentrations, clusters or other groupings and the
smooth ellipical morphology is typical of the stellar appearance of
Local Group dSph systems, particularly the Tucana
dwarf (Lavery \& Mighell \markcite{lm92} 1992).  It is possible to trace
the resolved component to a diameter of some 3 arcmin along the major axis
but as the majority of the flux resides in the unresolved component a better
approach for estimating the morphological structure is to analyze an
appropriately smoothed version of the image.  

\begin{table}
\begin{tabular}{lcr}
\tableline
\\
Cross-Identification &           & SGC 1001.9-2705 \\
	             &           & PGC 029194 \\
	             &           & AM 1001-270 \\
Distance             & $(m-M)_0$ & 25.3 $\pm$ 0.2 \\
	             & Mpc       & 1.15 $\pm$ 0.10 \\
Metallicity          & [Fe/H] & -1.6 $\pm$ 0.3 \\
Apparent Magnitude   & $m_{\rm V}$ & 14.8 $\pm$ 0.2 \\
                     & $m_{\rm R}$ & 14.3 $\pm$ 0.2 \\
Extinction           & E(B-V) & 0.05 \\
	             & $A_{\rm V}$ & 0.2 \\
Absolute Magnitude   & $M_{\rm V}$ & -10.7 $\pm$ 0.3 \\ 
Radial Profile       & $\Sigma_o$ (V mag arcsec$^{-2})$ & 24.3 $\pm$ 0.2 \\
(geometric mean      & r$_{c,g}$ (arcmin) & 0.80 $\pm$ 0.05 \\ 
parameters deduced   & r$_{{1 \over 2},g}$ & 0.73 $\pm$ 0.05 \\
from King model fit) & r$_{t,g}$          & 5.2  $\pm$ 0.2 \\
Ellipticity          & e & 0.35 $\pm$ 0.03 \\
	             & position angle & $145^\circ \pm 5$ \\
Position (B1950)     & RA & $10^h 01^m 47^s.5$ \\
	             & Dec & $-27^\circ 05' 15''$ \\
\\
\tableline
\\
\end{tabular}
\caption{Properties of the Antlia dwarf galaxy}
\label{props}
\end{table}

Prior to smoothing the individual broadband images two different methods
were investigated to reduce the influence of bright foreground stars, and the 
occasional background galaxy, superposed on the central parts of the CCD 
images.  The first method used a straightforward clipping algorithm.  All
pixels in the CCD frame were clipped to have an intensity no brighter than 
the peak intensity in any unambiguous Antlia stars.  This is straightforward
to accomplish since the tip of the Antlia population is at a well defined
magnitude (see Figures~\ref{cmd} and \ref{ilf}).  In the second method 
individual unsaturated stellar images were removed by point-spread-function
(PSF) subtraction of an average stellar profile, and saturated stellar images
or background galaxies were removed by excising a region around them.
Convolution with a flux-conserving Gaussian kernel was then used to obtain
a smooth image of the dwarf galaxy.  After this processing the ellipicity, 
radial profile, position angle of the  major axis and total integrated
flux relative to sky, were straightforward to obtain using standard image
analysis techniques (Irwin \markcite{i85} 1985).  These properties
together with other derived quantities discussed later are listed in
Table~\ref{props}.  Combining the apparent magnitudes with the distance
estimate derived below, yields an M$_{\rm V} = -10.7 \pm 0.3$ and an
underlying V--R color of $\approx$ 0.5.  These values are similar to the
brighter dSph galaxies of the Milky Way and very close to those for the
M31 dSphs And I, II, III.  For comparison, Lavery \& Mighell (1992) estimated
Tucana to have an absolute magnitude of M$_{\rm V} = -9.5$ and a B -- R color
$\approx 1.1$.

\subsection{Color-magnitude diagram}

\begin{figure}
\centerline{
\hbox{\psfig{figure=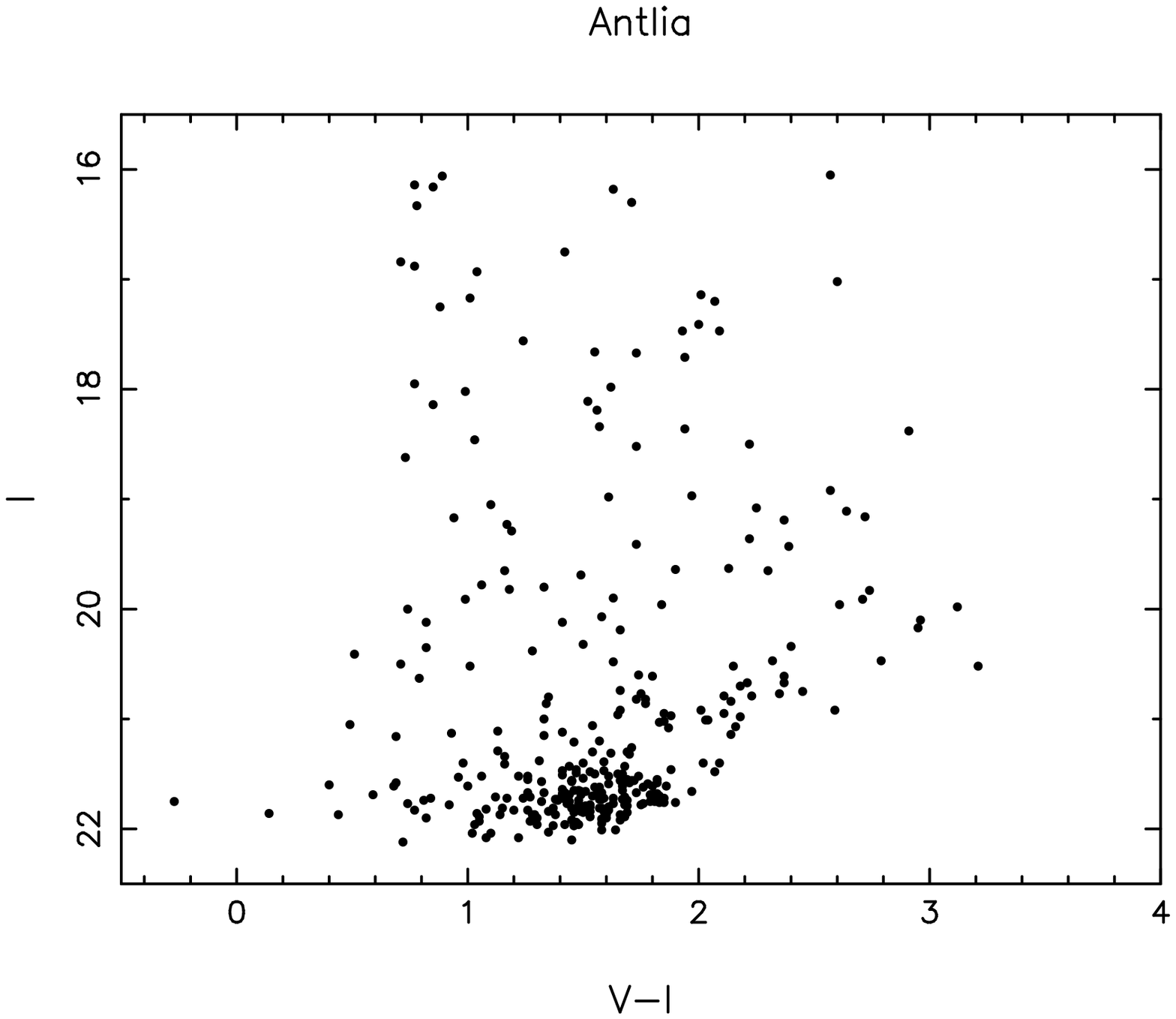,width=10cm,angle=-90}}
\hbox{\psfig{figure=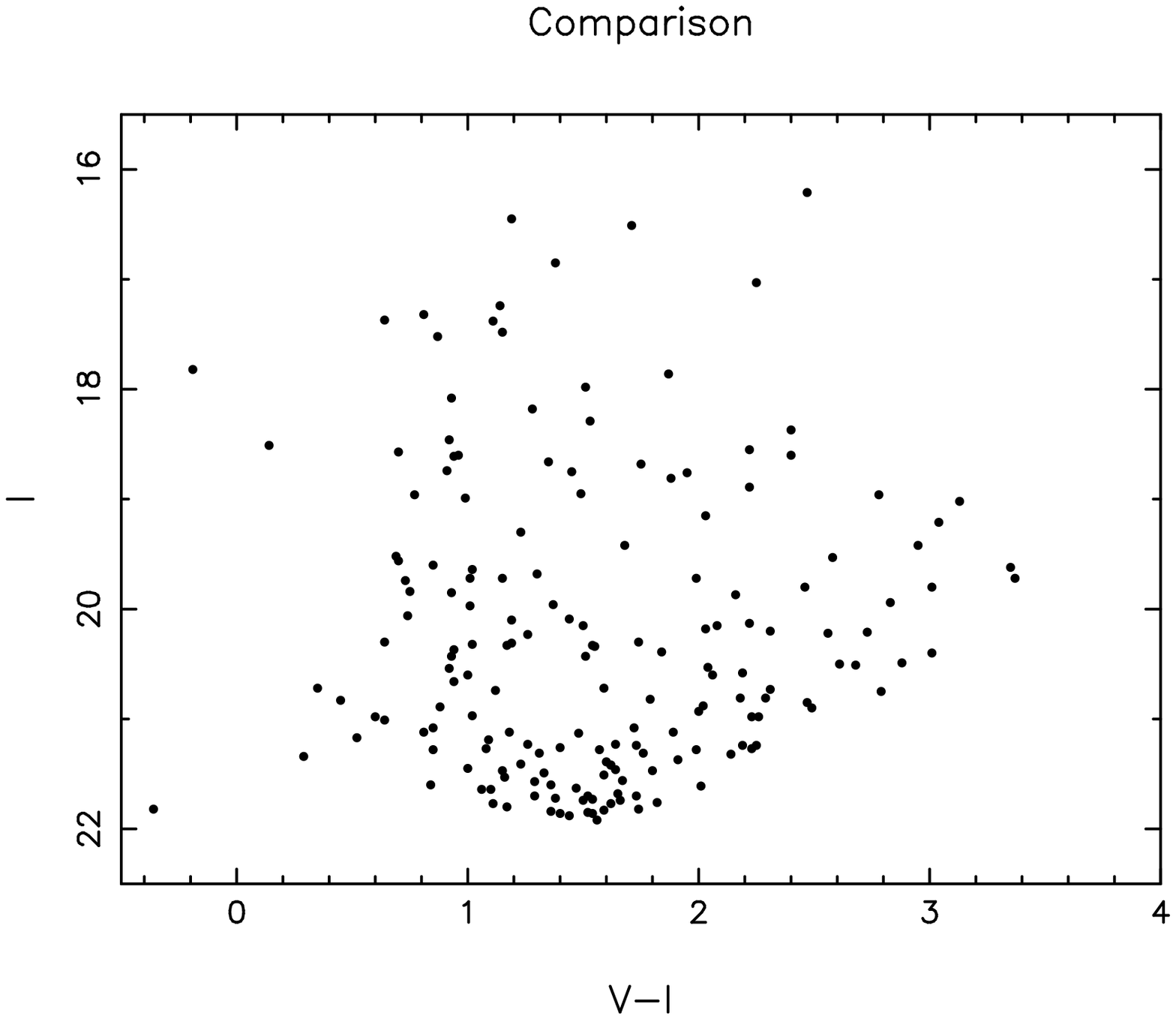,width=10cm,angle=-90}}
}
\label{cmd}
\end{figure}

The stacked V and I frames were analysed using a PSF fitting routine
and the derived intensities mapped on to a total intensity system
using aperture magnitues of isolated unsaturated bright stars (Irwin 1985).
A V,I color-magnitude diagram (figure~\ref{cmd}) constructed from a
4 arcmin diameter region centered on Antlia, compared to one constructed from
a similarly-sized region from the outer parts of the CCD frame, unambiguously
shows the tip of what appears to be a normal red giant branch; no bright blue 
stars from a young stellar population are evident in the galaxy, nor is there 
any compelling evidence for an extensive intermediate-age AGB component - 
although much deeper color-magnitude diagrams are needed to quantify this
latter assertion.  We tentatively conclude from figure~\ref{cmd} that the
stellar content of Antlia is most likely dominanted by an old stellar 
population similar to that found in most Local Group dSphs.

\subsection{Distance and metallicity}

\begin{figure}
\centerline{
\hbox{\psfig{figure=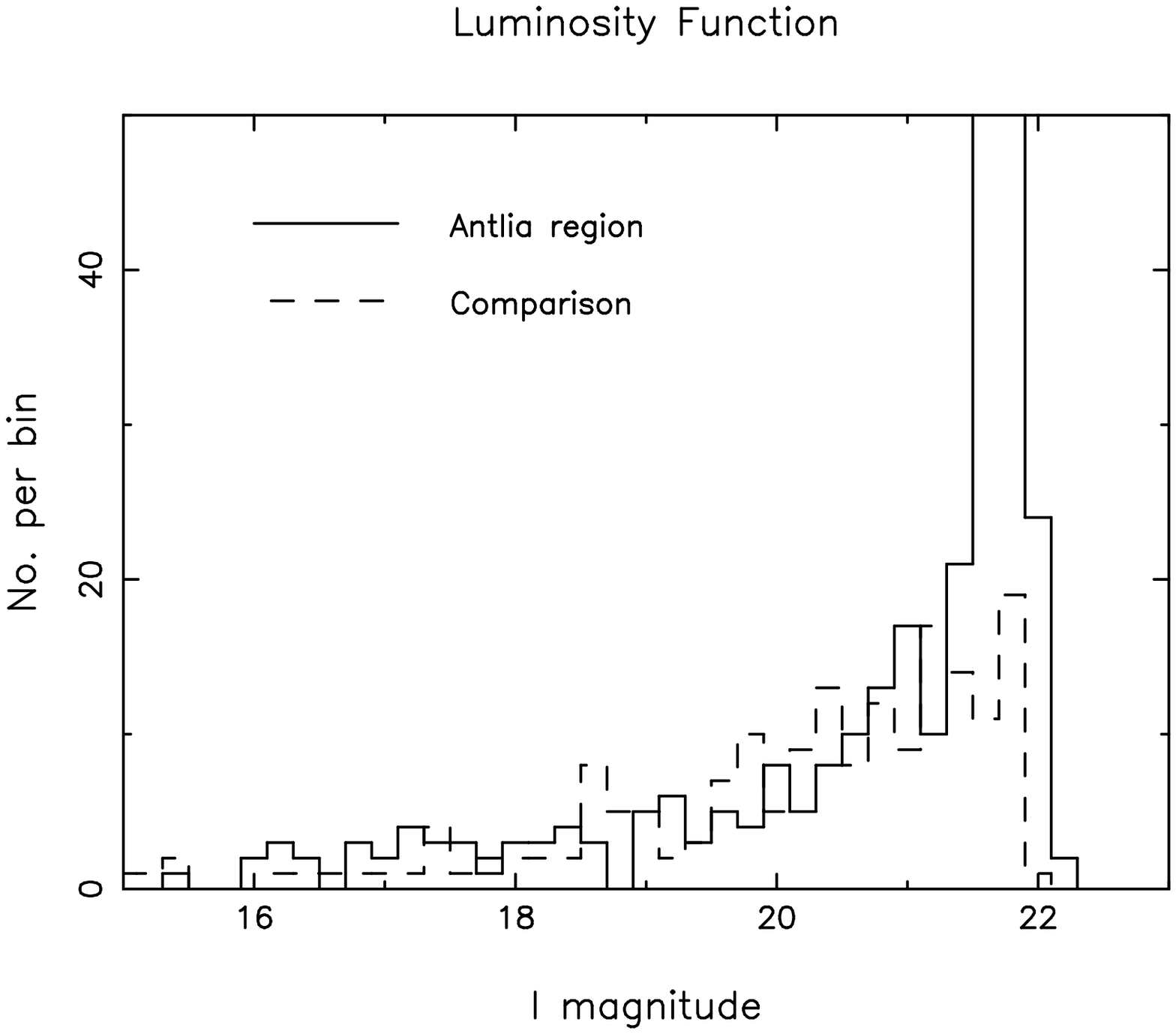,width=15cm,angle=-90}}
}
\label{ilf}
\end{figure}

We can use the well defined locus of the giant branch to provide a
provisional estimate of both the distance and metallicity of the dominant
stellar population (cf. Lee \markcite{l93}, 1993; Lee et al.\ \markcite{lal93}
1993).  The location of the tip of the giant branch is revealed more clearly
in the I-band luminosity functions presented in figure~\ref{ilf}. We estimate
I$_{\rm rgbt} = 21.4 \pm 0.1$ and from the V,I color distribution the 
average color at the tip of the giant branch is V--I $= 1.55 \pm 0.05$.  
Located at Galactic coordinates $l = 263.1, b = 22.3$, Antlia lies in a region
of relatively low Galactic foreground extinction.  From the maps of Burstein \&
Heiles (1982) we estimate the foreground extinction to be given by
E(B-V) $\approx$ 0.04-0.05 which translates to an approximate extinction in 
the I-band of A$_{\rm I} = 0.1$ and an a reddening of E(V-I) $= 0.06$ 
(cf.\ Irwin et al.\ \markcite{i95} 1995).  The expected absolute magnitude of
the giant branch tip is only weakly dependent on metallicity for the typical
metal poor populations of the Galactic dwarfs (Lee 1993), therefore taking
this absolute magnitude to be $-3.98 \pm 0.05$ gives a de-reddened distance
estimate for Antlia of (m - M)$_o = 25.3 \pm 0.2$, where the error includes
all the uncertainties above and an additional zero-point error from the
photometry of $\pm 0.04$.  

Following Lee et al. (1993) we can make a provisional estimate of the 
metallicity of the red giant branch population by comparing the locus of points
near the tip of the giant branch with other galaxies having similar published 
V,I color magnitude diagrams and with Galactic globular 
clusters.  With the caveat that we are not sampling much of the giant branch of
Antlia, both Leo I, with [Fe/H] $= -2.0 \pm 0.1$ (Lee et al. 1993), and 
Tucana, with [Fe/H] $= -1.8 \pm 0.2$ (Saviane et al. \markcite{sav96} 1996),
have giant branch loci slightly
blueward of Antlia after due allowance for reddening; while that of NGC 3109 
with [Fe/H] $= -1.6 \pm 0.2$ Lee et al. (1993) agrees to within the errors.  
An independent estimate can be derived from the mean metallicity--luminosity 
relation for dwarf galaxies (Da Costa et al.\ \markcite{dac91} 1991).  For an
M$_{\rm V} = -10.7$ dwarf the expected metallicity is [Fe/H] $\approx$ -1.7,
in reasonable agreement with the previous value.

\subsection{Radial profile}
\begin{figure}
\centerline{
\hbox{\psfig{figure=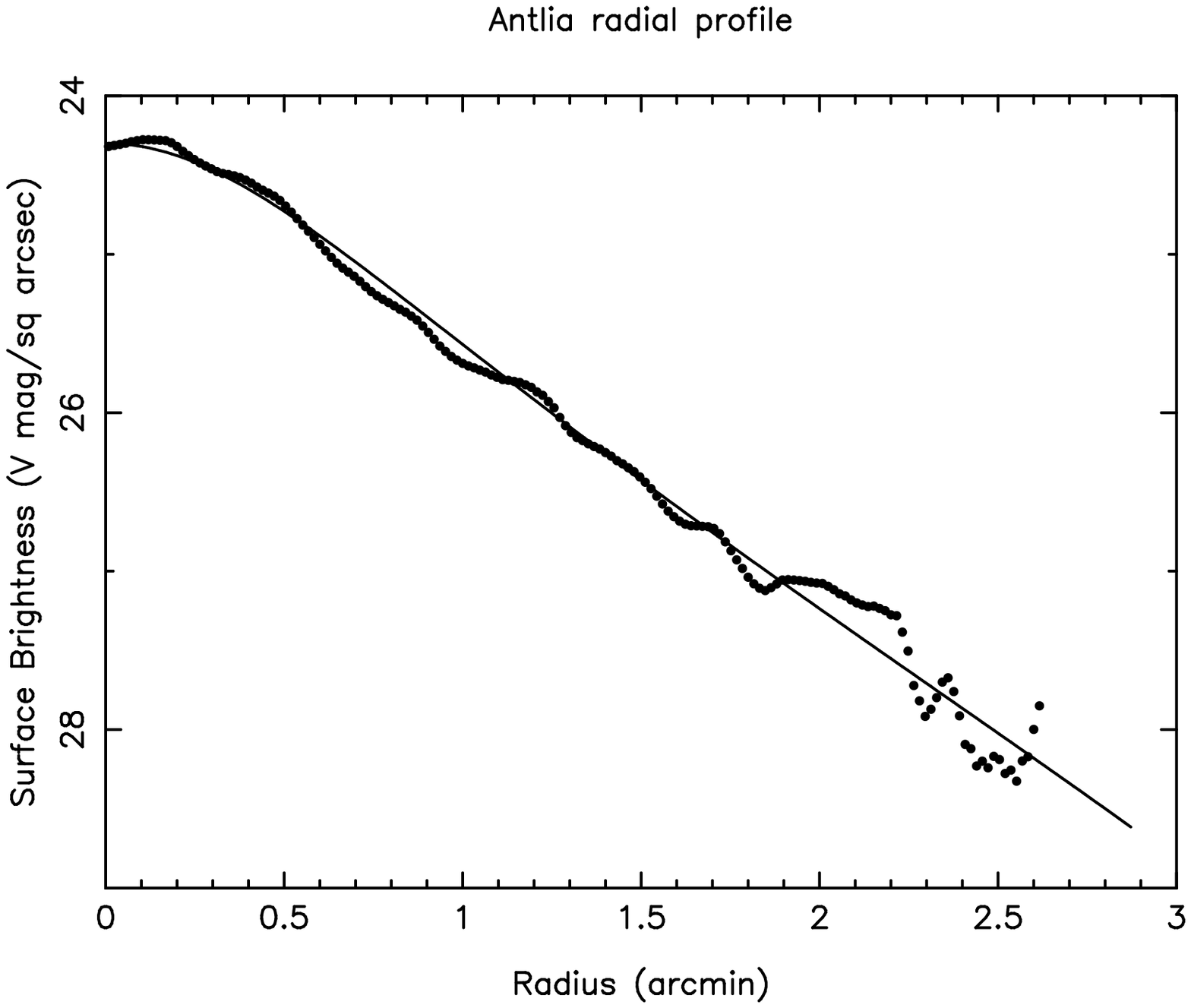,width=15cm,angle=-90}}
}
\label{rpf}
\end{figure}

The previously derived smooth images of the Antlia dwarf were also used to 
investigate the surface brightness profile.  The intensity-weighted centroid,
mean ellipticity and position angle were used to define concentric elliptical 
annulii and the average flux from the dwarf with respect to sky in these 
annulii recorded and used to produce the profile shown in figure~\ref{rpf}.  
A single component King Model (King \markcite{k62} 1962) 
fitted to the geometric mean of the semi-major and semi-minor axis profiles 
gives a convenient parameterisation of the profile and the results from the fit
are listed in Table~\ref{props} -- although we note that an exponential 
profile fits almost equally well.  The central surface brightness of 24.3 
$\pm$ 0.2 magnitude arcsec$^{-2}$ in the V-band, the core,
half-light, and ``tidal'' radii of 268 pc, 244 pc and 1739 pc respectively,
all lie comfortably within the range exhibited by the known Local Group
dwarf spheroidal galaxies (see, for example Irwin \& Hatzidimitriou 
\markcite{ih95} 1995).
The low concentration index, c $= 0.81$, and profile are most reminiscent
of Sextans and Sculptor both of which are also well fit by an exponential
profile.  

\section{Other Observations and Discussion}

The faint smudge that Antlia makes on the UKST sky survey plates
had been previously catalogued by Corwin et al. \markcite{cdvdv85} (1985);
Feitzinger \& Galinski \markcite{fg85} (1985); and Arp \& Madore
\markcite{am87} (1987).  Corwin et al. (1985) in particular noted it as a
possible nearby dwarf, but to our knowledge it has not previously been
followed up optically.  Curiously, in an HI survey of southern late-type 
galaxies, Fouqu\'e et al. \markcite{fbdgp90} (1990) detected Antlia at a 
heliocentric radial velocity of 361$\pm$2 km/s with a full velocity width at 
half maximum of 21$\pm$4 km/s and also suggested that Antlia was a candidate 
Local Group member.  Additional 21cm observations were made by Gallagher et 
al.\ \markcite{glm95} (1995) but they failed to detect Antlia and give an
upper flux limit of $<$150 mJy, which although not confirming the earlier 
result is still consistent with the peak flux of 122 mJy measured by 
Fouqu\'e et al.\ (1990).  Recently Thuan has been reported as independently
confirming the detection and velocity of Fouqu\'e et al.\ (Karachentsev
1997, private communication).  If the detection and integral HI parameters 
deduced by Fouqu\'e et al. are correct the total HI mass of Antlia is of
order 8$\pm$2 $\times 10^5 \Msun$, assuming small optical depth and a beamsize
larger than the extent of the dwarf (Roberts \markcite{r62} 1962).  Although
Antlia would be unique among dSph galaxies in having a significant HI mass,
this value would lie on an extension of the well defined correlation between
HI mass and B-band luminosity for LSB galaxies (Sprayberry et al. 
\markcite{spr95} 1995).

The velocity profile width derived by Fouqu\'e et al. (1990) can be
combined with the surface brightness properties to derive preliminary
total mass and mass-to-light ratio estimates for Antlia.  This is of
considerable interest since the apparent high mass-to-light ratios of the
Galactic dSphs could have been influenced by the presence of the Galactic
tidal field (Kuhn \markcite{kuhn93} 1993). Obtaining a similar result for
a more isolated dSph would help to shed light on this subject.  Assuming a
single component King model, the total mass of the system can be
calculated by applying the method of Illingworth \markcite{ill76} (1976). 
Using the parameters in Table~\ref{props} and deriving the line-of-sight
{\it rms} velocity, $\sigma_{rms} = 9 \pm 2$ km/s, from the HI profile of
Fouqu\'e et al. (1990), yields a total mass estimate of 3.3$\pm$1.3
$\times 10^7 \Msun$, where the error is dominated by the uncertainties in
the velocity dispersion.  The total mass-to-light ratio is then $20 \pm 8$
in solar units.  Alternatively we can use the method of Richstone \&
Tremaine \markcite{rt86} (1986), which is relatively insensitive to the
exact shape of the profile, to compute the central mass-to-light ratio. 
We find that Antlia has a central mass-to-light ratio $= 15 \pm 6$ , where
once again the error budget is dominated by the velocity profile
uncertainty.  Further HI observations are urgently needed to improve the
accuracy of these estimates, since the derived mass-to-light ratios are
tantilisingly close to those expected by comparison with similarly
luminous Galactic dSphs. 

Optically, Antlia appears to be a normal dwarf spheroidal in that it has
no young stars or HII regions, a smooth morphology and a color magnitude
diagram similar to the rest of the Galactic dSphs.  As such we would have
expected it to have little or no gas, so the HI measurement by Fouqu\'e et al.
(1990) is very interesting.  Galactic satellite dSphs may have had their
gas tidally stripped by repeated interactions with the host system and it
has been suggested (for instance, by
Faber \& Lin \markcite{fb82} 1982) that
their precursors could be the numerous dwarf irregular systems found in
isolated regions in the Local Group.  If Antlia is an otherwise typical
dSph, but is confirmed to have gas, then it would be a unique galaxy in the
Local Group.  Of the other comparable outlying dwarves, LGS3 and Phoenix
have populations of young stars and detected gas, while Tucana appears to
be a gasless dSph.

\begin{figure}
\centerline{\hbox{\psfig{figure=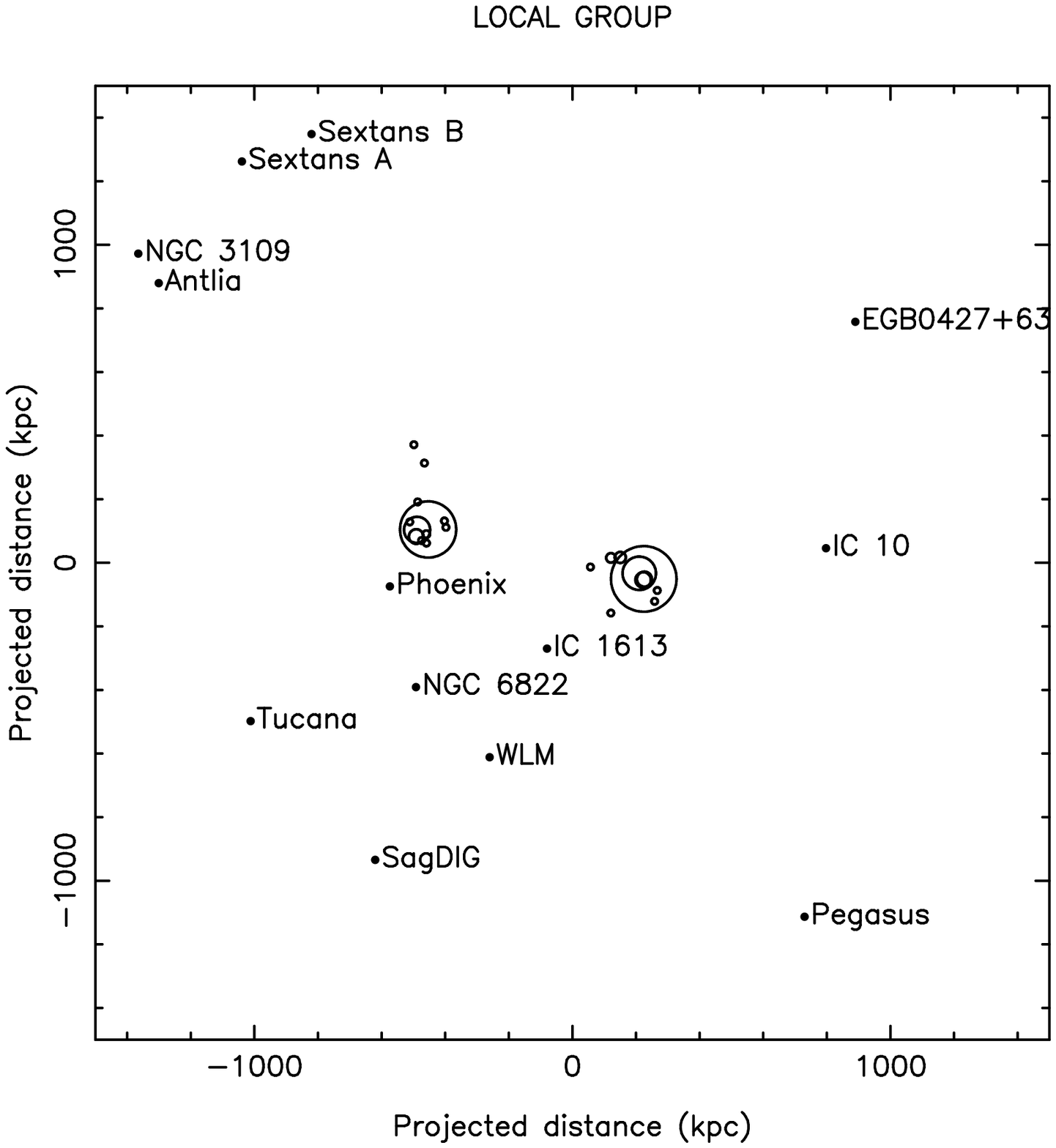,width=20cm,angle=-90}}
}
\label{localgroup}
\end{figure}

At a Galactocentric distance of 1.15 $\pm 0.10$ Mpc, Antlia lies close to 
the boundary of the Local Group and interestingly is only 1.2 degrees away on 
the sky from the Local Group dwarf NGC 3109 and at a comparable distance 
(see figure \ref{localgroup}).  Assuming that the HI velocity is correct,
Antlia has a velocity of $\approx$78 km/s with respect to the center-of-mass
of the Local Group, compared to $\approx$120 km/s for NGC3109, and as such
both systems are almost certainly bound to the Local Group.  Their proximity 
suggests that they may themselves form a bound pair.
 
At the limit of their observed
rotation curve, Jobin \& Carignan \markcite{jc90} (1990) show a total mass 
for NGC 3109 of between 8.6 and 8.9 $\times 10^9 \Msun$.  From the apparent
separation on the sky we know that the minimum possible distance between 
Antlia and NGC 3109 is $\approx$26 kpc. With these best case parameters 
the total velocity difference would have to be less than about 54 km/s for 
Antlia to be bound. On purely gravitational criteria, assuming the total
mass of the Local Group is $\approx 3 \times 10^{12} \Msun$, the tidal field
of the Local Group exceeds that of NGC3109 if Antlia is more than $\approx$
130 kpc from NGC3109 -- a weaker constraint than the previous one.  
Therefore we can not yet rule out the possibility of Antlia being a satellite
of NGC3109, confirmation, or otherwise, awaits supportive HI measurements for
Antlia and a good differential estimate of the distance between NGC3109
and Antlia.

Further optical and radio observations are urgently required to unlock the
secrets of this fascinating system.

\acknowledgements{The authors wish to thank John Pilkington of the
Royal Greenwich Observatory for electronically scanning many survey
plates; Sue Tritton of the Royal Observatory Edinburgh for her
help in distinguishing faint astronomical objects from plate defects;
and I. Karachentsev for his valuable insights and assistance.
This research has made use of the NASA/IPAC Extragalactic Database (NED)
which is operated by the Jet Propulsion Laboratory, California Institute
of Technology, under contract with the National Aeronautics and Space
Administration.}

\newpage
\figcaption[whiting.fig1.ps]
{\footnotesize Number counts of galaxies with v $<$ 500 km/s.  The
Local Group contributes almost all of the galaxies below M$_{\rm B} = -11$.
Data from Schmidt \& Boller 1992.}

\figcaption[whiting.fig2.ps]
{\footnotesize The new Local Group dwarf galaxy
Antlia.  CTIO 1.5m telescope images in V, R and I with total exposure times of
4800s, 3600s and 3600s respectively were combined to produce this ``true''
color picture.  The field shown covers 4 $\times$ 4 arcmin of sky with
North to the top and West to the left.}

\figcaption[whiting.fig3a.ps]
{\footnotesize Left: V,I color-magnitude diagram for Antlia, showing
the tip of the red giant branch; right, the same diagram for the outer parts
of the same CCD frames showing the lack of the red giant population.}

\figcaption[whiting.fig4.ps]
{\footnotesize The I-band luminosity function of a region centered on
Antlia (full line) compared with the local field luminosity function
(dashed line).  The tip of the red giant branch at I $=$ 21.4 is very clearly
defined.}

\figcaption[whiting.fig5.ps]
{\footnotesize The geometric mean of the semi-major and semi-minor
V-band surface brightness profiles of Antlia.  A single component,
three-parameter King model fit is shown overlain, although a two-component
exponential profile fits almost equally well.}

\figcaption[whiting.fig6.ps]
{\footnotesize Projection of the galaxies in the Local Group
onto a convenient plane.  The large circle at left is the Milky
Way, surrounded by its satellite galaxies; at right is M31 with its
satellites.  Antlia is located near NGC 3109, but distant from almost
all other dwarves on the outskirts of the Group.  Data taken from
Hodge (1994).}

\end{document}